\documentclass[pdflatex,sn-mathphys-num]{sn-jnl}% Math and Physical Sciences Numbered 

\usepackage{graphicx}%
\usepackage{multirow}%
\usepackage{amsmath,amssymb,amsfonts}%
\usepackage{amsthm}%
\usepackage{mathrsfs}%
\usepackage[title]{appendix}%
\usepackage{xcolor}%
\usepackage{textcomp}%
\usepackage{manyfoot}%
\usepackage{booktabs}%
\usepackage{algorithm}%
\usepackage{algorithmicx}%
\usepackage{algpseudocode}%
\usepackage{listings}%

\usepackage{graphicx}% Include figure files
\usepackage{dcolumn}% Align table columns on decimal point
\usepackage{bm}% bold math
\begin{document}

%\title{Stabilization of a micro-resonator Brillouin laser to the $5S_{1/2}\rightarrow 5D_{5/2}$ two-photon transition in rubidium}% Force line breaks with \\

\title[A narrow-linewidth Brillouin laser for a two-photon rubidium frequency standard]{A narrow-linewidth Brillouin laser for a two-photon rubidium frequency standard}

\author[1]{\fnm{Kyle W.} \sur{Martin}}%\email{kyle.martin@avinc.com}

\author[1]{\fnm{River} \sur{Beard}}%\email{river.beard@avinc.com}

\author[2]{\fnm{Andrei} \sur{Isichenko}}

\author[2]{\fnm{Kaikai} \sur{Liu}}

\author[3]{\fnm{Seth E.} \sur{Erickson}}

\author[3]{\fnm{Kaleb} \sur{Campbell}}
\author[2]{\fnm{Daniel J.} \sur{Blumenthal}}
\author*[3]{\fnm{Sean} \sur{Krzyzewski}}\email{qst@afrl.af.mil}

\affil[1]{\orgname{AV Incorporated}, \orgaddress{\street{1300 Britt Street}, \city{Albuquerque}, \postcode{87123}, \state{NM}, \country{USA}}}

\affil[2]{\orgdiv{Department of Electrical and Computer Engineering}, \orgname{University of California}, \orgaddress{\city{Santa Barbara}, \postcode{93106}, \state{CA}, \country{USA}}}

\affil*[3]{\orgdiv{Space Vehicles Directorate}, \orgname{Air Force Research Laboratory}, \orgaddress{\city{Kirtland Air Force Base}, \postcode{87117}, \state{NM}, \country{USA}}}

\date{\today}% It is always \today, today,

\abstract{\unboldmath High precision portable and deployable frequency standards are required for modern navigation and communication technologies. Optical frequency standards are attractive for their improved stability over their microwave counterparts; however, increased complexities have anchored them in the laboratory. Sacrificing sensitivity of the most stable optical clocks has led to the recent development of deployable and portable optical frequency standards, leveraging hot atomic or molecular vapor. The short term limit for a majority of previous reports on two-photon rubidium standards is either the shot-noise or intermodulation limit hindering the one second fractional frequency stability to around $1\times10^{-13}/\sqrt{\tau}$. The answer for the shot-noise limit is to increase optical power and collected fluorescence, while the intermodulation limit solution requires improvements in laser linewidth, stimulated Brillouin scattering (SBS) lasers are known to reduce frequency noise, suppressing noise of the pump laser at high offset frequencies.  We investigate an optical frequency standard based on the two-photon transition in $^{87}$Rb probed with a narrow linewidth photonic integrated circuit SBS laser with a quality factor over 130 million and instantaneous linewidth $<$ 10 Hz. The use of a narrow linewidth clock laser coupled with operating at higher optical intensities yields clock instabilities of $2\times10^{-14}$ at one second, currently the best reported short-term stability for a two-photon rubidium optical frequency standard. }

\maketitle

%\tableofcontents
%\linenumbers
\section{\label{sec:intro}Introduction}

Modern communication, navigation, and sensing systems rely on portable clocks and oscillators to provide necessary stable frequency and phase sources \cite{Maleki_2005}. While crystal and MEMS oscillators meet a vast majority of timing needs, more precise and robust applications require atomic clocks. Microwave clocks based on rubidium and cesium are currently the most popular technology fulfilling this need \cite{Maleki_2005, McNeff2002}.  These atomic microwave standards have provided robust and stable phase and frequency sources and have well understood frequency shifts and drifts \cite{Formichella2017}. Nonetheless, emerging technologies such as femtosecond two-way optical time transfer \cite{Sinclair2019, Hugo2016, Deschenes2016, Isaac2018, Bigelow2019, Olson2025} and high frequency distributed antenna systems \cite{Holtom2024,Mudumbai2009} require even more stable low size, weight, and power (SWaP) frequency standards.  These emerging technologies coupled with improvements in existing communication, navigation, and sensing technologies adds significant value to portable atomic frequency standards with superior timing stability.   

Current state-of-the-art single ion \cite{Huntemann2016,Marshall2025,Brewer2019} and optical lattice clocks \cite{Grebing:16,Bothwell_2019,McGrew2018} yield excellent results in controlled laboratory environments. Efforts to bring these more advanced optical lattice and ion clocks from a laboratory environment into the field is an active area of research \cite{Delehaye30032018,Kong2020ATO,Takamoto2020TestOG,Ohmae2021TransportableSO,Origlia2018TowardsAO,Khabarova2022TowardAN}.  However, single ion and optical lattice clocks have yet to make an impact with more stringent requirements of  SWaP, robustness, and autonomy necessary for widespread communication, navigation and sensing.  Selective tradeoffs need to be made for advanced deployable clocks to be used in a more widespread manner.  

%Additionally, the advent of the vibration-immune fiber frequency comb \cite{Sinclair2015} has become a driving force for further innovation in field-deployable optical clocks. 

 In order to meet the stringent demands for deployable clocks, research teams have often elected to simplify the state-of-the-art optical clocks to prioritize certain advancements in frequency stabilization. Several efforts have maintained a fundamental microwave clock frequency while adding lasers for state preparation \cite{Gozzelino2023} or for laser cooling to reduce Doppler and collisional shifts \cite{Ascarrunz2018}. Nevertheless, these microwave standards still lack the higher short term stability of their optical counterparts.  Portable optical cavities have also garnered interest in recent years \cite{Cole:24, Kelleher:23}; while generally superior in performance at short-time scales, they lack an atomic reference and suffer from long-term frequency drifts.  The use of hot atomic \cite{Hilton2025} or molecular vapors \cite{Roslund2024} have had recent success operating outside the laboratory in a relevant SWaP envelope. These hot atomic or molecular optical clocks retain high stability at short timescales (leveraging stable optical oscillators) while simultaneously benefiting from long-term stability of the probed atomic or molecular energy transitions. 

We investigate the optical rubidium atomic frequency standard leveraging the $5S_{1/2}(F=2) \rightarrow 5D_{5/2} (F=4)$ transition  in $^{87}$~Rb and detecting fluorescence decay along the $6P_{3/2}\rightarrow5S_{1/2}$ \cite{Martin2009,2photonreview,Erickson:24,Maurice:20,Newman:19,Callejo:25,Li2024,Beard:24,Lemke2022,POULIN2002,Gerginov2018,Perrella2019}.  This transition has several advantages as the required clock laser can be generated through second harmonic generation of telecom C-band laser systems, the fluorescence is easy to distinguish from the clock laser at 778~nm, as well as the relatively high excitation rate for a two-photon system.  However, the short term clock performance is generally limited by either photon shot-noise or laser frequency noise in the form of intermodulation limit \cite{2photonreview}. In this article, we demonstrate an order-of-magnitude improvement on both of the short-term limitations by using high intensity to suppress photon shot-noise and a photonic integrated silicon nitride stimulated Brillouin scattering (SBS) laser to reduce the intermodulation limit.  

%mention that molecular iodine uses a modulation transfer technique technique measuring absorption, they can modulate faster and have high SNR since they operate near resonance.  Fundamentally, molecular iodine for a give optical intensity has lower shot noise limit and intermodulation limit.  

\section{\label{sec:short} Short-term clock limits}

\begin{figure}
\includegraphics[width=\columnwidth]{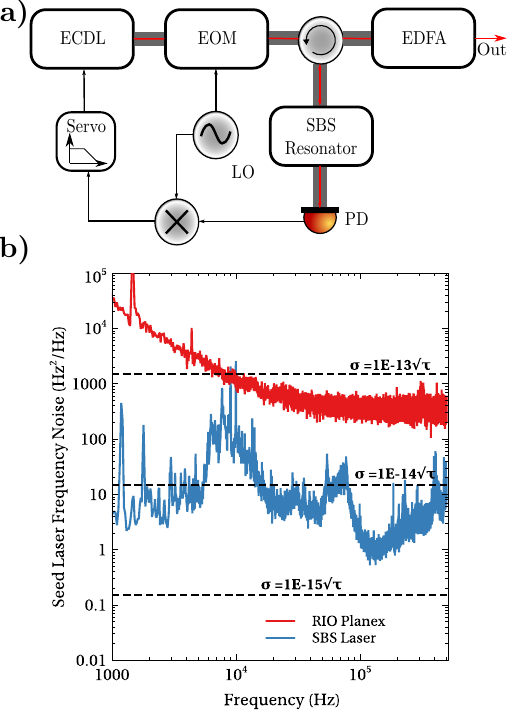}% Here is how to import EPS art
\caption{\label{fig:experimental_SBS} (a) System diagram of the ECDL locked to an integrated silicon nitride resonator, as described in the text.  The stimulated Brillouin scattering output (labeled out) is used to probe the $5S_{1/2}\rightarrow 5D_{5/2}$ two-photon transition. (b) The decrease in measured frequency noise of the SBS laser over the ECDL as measured against a stable cavity locked laser (fractional frequency $<3\times 10^{-15}$ at one second).  Horizontal lines indicating intermodulation limited clock performance are also shown.  PD-photodiode, EDFA- erbium doped fiber amplifier, EOM- electro-optic modulator, ECDL- external cavity diode laser, LO-local oscillator, SBS-stimulated Brillouin scattering.}
\end{figure}

%The major difference in short-term performance between molecular iodine and two-photon rubidium remains the intermodulation limit.  Since modulation transfer spectroscopy is fundamentally a absorption measurement, whose modulation frequency is not limited by the lifetime of the atoms larger modulation frequencies can be used thus reducing the intermodulation limit of the frequency standard, as the frequency noise of the laser typically falls with increasing offset frequencies.  The lifetime of the $5D_{5/2}$ state of rubidium, measured to be 238.2 (2.3)~ns \cite{Sheng2008}, typically limits modulation frequencies to $<$~1~MHz, increasing restrictions on laser frequency noise at small offset frequencies.  Reducing laser frequency noise at twice the working modulation frequency could increase the short term stability of the two-photon rubidium optical frequency standard.  
% (for degenerate two-photon transitions this is twice the laser frequency)
%A frequency standard stability is normally measured and reported using the fractional frequency Allan deviation $\sigma_y$ \cite{Allan}.  

A recent review article of the $5S_{1/2}\rightarrow 5D_{5/2}$ two-photon transition in Rb \cite{2photonreview} concluded that for the majority of reports the short-term limit is either the photon shot-noise or laser frequency noise in the form the the intermodulation limit.

\subsubsection{Photon Shot-Noise Limit}
  The statistical uncertainty involved in detecting photons gives rise to a fundamental limitation to any photo-detector known as photon shot-noise. The photon shot-noise sets the limit for references \cite{Ahern2025,Li2024,Beard:24,Lemke2022,Martin2019,POULIN2002,Gerginov2018,Perrella2019}. A frequency standard whose Allan deviation \cite{Allan} is limited by photon shot-noise limit can be written as \cite{Hilico1998}, 
\begin{equation} \label{eq:SN1}
    \sigma^{(SN)}_y = \frac{1}{\nu_0}\sqrt{\frac{S_f}{2\tau}},
\end{equation}
where,
\begin{equation}\label{eq:SN2}
    S_f = \Bigg(\frac{g}{p}\Bigg)^2\frac{S_v}{2}.
\end{equation}
$\tau$ is the averaging time, $\nu_0$ is the clock transition frequency, $g$ is the mixer gain, $p$ is the error signal slope and $S_v$ and $S_f$ are the measured voltage and frequency noise, respectively, in the detector bandwidth.    Reduction of shot-noise can be achieved by improving the error signal slope, $p$.  In fluorescence based frequency standards $p$ can be increased by improving collection efficiency, although this has a upper limit. The collection efficiency for this work and Lemke et al. \cite{Lemke2022} is $\sim 7-8\%$ of the fluoresced photons, with the currently best reported number Perrella of $15\%$ \cite{Perrella_2022} Alternatively, amplified  probe light intensity can increase $p$, consequently, ac-Stark effects may limit long-term performance \cite{Martin2019}.

\subsubsection{Intermodulation Limit}
All oscillators have broadband noise, consequently, this noise bypasses filtering and demodulation in error signal generation.  In continuous wave frequency modulation this is known as the intermodulation limit for the oscillator (the Dick effect for pulsed systems). The intermodulation limit sets the short-term performance for references \cite{Maurice:20,Newman:19}, restricting fractional frequency performance to $3 \times 10^{-12}$ at one second. The intermodulation limit is derived in \cite{Audoin1991} and can be written as, 
\begin{equation} \label{eq:IM}
    \sigma^{(IM)}_y = \frac{(S^{(LO)}_y [2f_m])^{1/2}}{2\sqrt{\tau}},
\end{equation}
where $S^{(LO)}_y [2f_m]$ is the laser frequency noise at twice the modulation frequency.  Ahern et al. \cite{Ahern2025} noted that for multi-photon processes the frequency noise on each photon needs to be included. Consequently, the total state intermodulation limit needs to sum over the frequency noise of each excitation photon, this was overlooked in Martin et al. \cite{Martin2019}. Correlated noise from two sources ($e_1$ and $e_2)$ adds as,
\begin{equation}
    e_y = \sqrt{e_1^2 +e_2^2+2ce_1e_2},
\end{equation}
where $-1\le c \le 1$ is the correlation parameter.  The intermodulation limit for a degenerate two-photon excitation ($c=1$), can now be re-written, 
\begin{equation} \label{eq:IM2}
    \sigma^{(IM)}_y = \frac{(S^{(LO)}_y [2f_m])^{1/2}}{\sqrt{\tau}}.
\end{equation}
%whose atomic interaction occurs on a scale much shorter than the laser coherence length,
  Typically, laser noise is inversely proportional to offset frequency.  Consequently, the intermodulation limit can be suppressed by increasing the modulation frequency, although the atomic lifetime sets an upper limit for this noise reduction technique for florescence detection, for two-photon rubidium this upper limit is $\sim 2$~MHz.  Ultimately, for frequency standards operating in the intermodulation limit a higher quality, narrower laser is required.

%In our system both the photons originate from the same optical source.  Since, atomic interaction occurs on a scale much shorter than the laser coherence length (which is $>1$~km), the intermodulation limits arising from each laser are correlated with each other.  
%Reducing the intermodulation limit of the system requires reducing the frequency noise of the excitation laser. This can be achieved by increasing the modulation frequency, since lasers frequency noise is generally inversely proportional to offset frequency, with the maximum modulation frequency limited by the excited state lifetime for fluorescence based techniques. 

In some instances both the photon shot-noise and intermodulation limits are sufficiently small and the short-term limit is driven by environmental instabilities.   This is the case in Perrella et al. \cite{Perrella:13}, where clock laser alignment limited performance. Reducing the short-term limit of the two-photon Rb frequency standard requires reduction of both the intermodulation and shot-noise limits as well as careful consideration to all environmental instabilities.

%When the instabilities are uncorrelated, the limit of a frequency standard is the quadrature sum of all instabilities at a specific timescale.  

\section{\label{sec:Exp} Materials and Methods}

\begin{figure}
\includegraphics[width=\columnwidth]{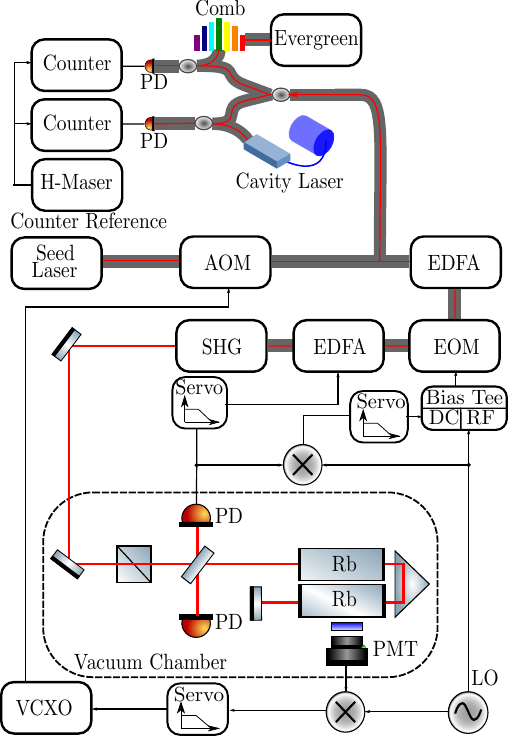}% Here is how to import EPS art
\caption{\label{fig:experimental_atoms}System diagram of the SBS laser locked to the two-photon transition in Rb as described in the text.  The clock laser is directly compared via a heterodyne with both a cavity-stabilized laser and an optical frequency comb stabilized to a molecular iodine standard labeled Evergreen in the figure. PD-photodiode, EDFA- erbium doped fiber amplifier, EOM- electro-optic modulator, LO-local oscillator, SBS-stimulated Brillouin scattering, VCXO-voltage controlled crystal oscillator, AOM -Acoustic optic modulator, SHG-second harmonic generation, PMT-Photomultiplier Tube.}
\end{figure}

The experimental apparatus is evolved from the system described in \cite{Lemke2022}.  However, in order to suppress both intermodulation and shot-noise the experiment operates at increased optical intensities, relative to \cite{Lemke2022}, while reducing laser frequency noise by stabilizing the clock laser to a stimulated Brillouin scattering (SBS) resonator \cite{Gundavarapu2019}. The SBS laser resonator is fabricated on an ultra-low-loss silicon nitride platform \cite{8472140}. The waveguide dimensions are $40$~nm in thickness and $11$~$\mu$m in width, designed to achieve low propagation loss and a low Brillouin lasing threshold. The radius of the resonator is 11.787~mm, ensuring the resonances match the Stokes shift \cite{Puckett2021}. The SBS laser has reduced laser phase noise in the high-frequency offset region ($<$~$100$~kHz), propagation loss of $0.1$~dB/m, and the loaded quality factor (Q) is $130$ million at $1550$~nm. These SBS lasers can reach a low lasing threshold of $14$~mW and a sub-Hz fundamental linewidth \cite{Gundavarapu2019}. Stabilizing the SBS output to the $5S_{1/2}\rightarrow 5D_{5/2}$ two-photon transition in $^{87}$Rb similar to Isichenko et al. \cite{Isichenko2024} will improve SBS stability at timescales greater than 1 second.

%where the electronic system bandwidth can limit the noise reduction of stabilized lasers. Fabrication details can be found in a previous publication .

%We aim to improve the stability of the SBS laser at timescales greater than 1 second by locking to the $5S_{1/2}\rightarrow 5D_{5/2}$ transition in $^{87}$Rb similar to reference \cite{Isichenko2024}.

Two separate experiments with different clock lasers were performed: one utilizing a narrow 1556~nm RIO Planex external cavity diode laser (ECDL) with a measure instantaneous  linewidth $1.1(15)$~kHz producing 20~mW of optical power, the second experiment stabilizes the same ECDL to a resonator generating SBS, see Figure \ref{fig:experimental_SBS}. To produce SBS, light from the ECDL is sent through a fiber coupled waveguide electro-optical phase modulator (EOM) driven at $22$~MHz to generate frequency sidebands for Pound-Drever-Hall (PDH) lock \cite{DREVER1983}.  The light is then sent through a circulator to the integrated resonator, where the transmitted light is detected using a InGaAs photodiode. The counter-propagating SBS first order Stokes signal is sent back through the circulator and then to an erbium doped fiber amplifier (EDFA).  After this point the laser system for the two separate experiments are identical.  Figure \ref{fig:experimental_SBS} also shows the measured decrease in frequency noise of the SBS output over the ECDL, measured directly via optical heterodyne between the seed laser and a reference cavity laser with a sub-hertz linewidth, SLS-INT-1550-200-1 \cite{SLSDATA}.

%The output of the detector is mixed with a copy of the 22~MHz EOM drive signal generating an error signal used to PDH lock the pump ECDL laser to the micro-resonator. 

%The laser, ECDL or SBS, now needs to by amplified, modulated, frequency doubled, and stabilized to the atomic transition, see Figure \ref{fig:experimental_atoms}. 
Figure \ref{fig:experimental_atoms} shows the experimental diagram after seed laser generation.  The laser, ECDL or SBS, is sent to a fiber coupled acoustic optical modulator (AOM) frequency shifter, the actuator for atomic stabilization.  At this point approximately ten percent of the optical power is sampled to be measured against a reference. The laser is sent through a pre-amplification EDFA, followed by another  low RAM EOM \cite{MartinPHD,Zhang2014} modulated at 101~kHz, targeting the lowest phase noise of the SBS and ECDL. The light is sent through another EDFA  and undergoes second harmonic generation (SHG) in a PPLN waveguide producing as much as 100~mW of light at 778.1~nm.  

%this time formed in a temperature stabilized proton exchange waveguide in lithium niobate, The choice EOM and decision to temperature stabilize were made to reduce nominal residual amplitude modulation (RAM) levels, as proton exchange EOMs typically have less RAM \cite{MartinPHD} and temperature fluctuations are a known source of quadrature RAM \cite{Zhang2014}.At this point we observed intense green light not confined to a single spacial mode, third harmonic generation from the PPLN  .

The generated 778.1 nm light is carried by optical fiber to a vacuum chamber enclosing the Rb cell apparatus. The entire free-space portion is mounted to a temperature controlled water-cooled breadboard. The light is collimated to a beam with an intensity radius ($1/e^2$) of $w_0=$~1.05(2)~mm.  The free space laser is directed through a long-pass filter to eliminate unintended third harmonic light \cite{Zhang:23}, a polarizer, and a pellicle beam-splitter to sample the light for power stabilization.  The beam passes through two fused silica vapor cells (length 50 mm, diameter 10 mm) connected via a 180$^{\circ}$ turning prism and enclosed in mu-metal to reduce background magnetic fields. The cells are filled with isotopically pure $^{87}$Rb and heated to 100$^{\circ}$C using low field resistive heaters.  Although previous experiments used a cat's eye retro-reflector \cite{Martin2018, Lemke2022, Beard:24}, a flat mirror was chosen to reduce pointing stability requirements on the collimator \cite{Martin2019}. A large area PMT is mounted on top of the vapor cells behind a short-pass optical filter to collect the 420~nm fluorescence from the rubidium atoms with high efficiency.  

The PMT output is demodulated with a temperature stabilized analog mixer at 101~kHz to generate the error signal. An analog servo controller with dual integrators and approximately 20 kHz bandwidth stabilizes the laser frequency by steering the shift applied by the AOM to the peak of the fluorescence signal.

InGaAs photodiodes  with low capacitance ($< 2$~nF) in photovoltaic mode detect each reflection from the pellicle beam-splitter. Although the responsivity is lower for InGaAs than Si, the differential temperature responsivity coefficient for InGaAs is smaller than Si, making the detector less susceptible to thermal fluctuations.  The detector response is amplified using a transimpedence amplifier (bandwidth $>1$~MHz); the 778.1~nm laser power is stabilized to the forward beam pick-off adjusting the pump current of the final EDFA.  Additionally, the signal from the photodiode is mixed with 101~kHz sinusoidal signal generating an error signal used to suppress RAM by feedback to a dc-bias across the EOM \cite{Zhang2014}.   The second photodiode is used as a witness for both the power servo and RAM servo, used to measure how well these environmental instabilities are suppressed.

The stabilized 1556~nm seed light (ten percent sampled after the AOM), is then sent to be compared against two references: the SLS cavity laser and a Vector Atomic Evergreen-30 iodine frequency standard, via optical heterodyne.   Each heterodyne detection was measured with a frequency counter referenced to a hydrogen maser.  Since the heterodyne is generated in the optical domain, the frequency counter only requires mHz level measurement stability to measure optical frequency instabilities $<1\times 10^{-15}$.

\section{\label{sec:results}Results}

%\subsection{Short-Term Stability}
%Environmental parameters known to impact clock performance were measured and properly scaled using Refs. \cite{Martin2018, Martin2019}.  

Measured clock performance can be impacted by environmental variables, reference clock performance, shot-noise and intermodulation noise.  Ideally, all of the environmental impacts on the frequency standard are well understood, measured, and suppressed.  However, equation \ref{eq:adev} sums the intermodulation, shot-noise limit, and direct measurements of remaining impacts from ac-Stark ($\sigma_y^{(ac-Stark)}$), RAM ($\sigma_y^{(RAM)}$), Rb-Rb collisional $\sigma_y^{(Rb-Rb)}$, and the reference clock ($\sigma_y^{(ref)}$), yielding a prediction of clock performance from environmental effects,
%\begin{widetext}

\begin{equation}
\begin{split}
\sigma_y(\tau)^2 & = (\sigma_y^{(IM)}(\tau))^2+(\sigma_y^{(SN)}(\tau))^2 \\\\
& +(\sigma_y^{(RAM)}(\tau))^2+(\sigma_y^{(ac-Stark)}(\tau))^2+(\sigma_y^{(Rb-Rb)}(\tau))^2+(\sigma_y^{(ref)}(\tau))^2.
    \label{eq:adev}
    \end{split}
\end{equation}
%\end{widetext}

\begin{table}%The best place to locate the table environment is directly after its first reference in text
\caption{\label{tab:table1}%
Free-space laser measured parameters alongside measurements necessary to calculate the intermodulation and shot-noise limit of a given experiment.
}
%\begin{ruledtabular}
\begin{tabular}{{@{}ll@{}}}
\multicolumn{2}{l}{\textbf{Shared values}}\\
\midrule
\textrm{Parameter}&
\textrm{value}\\
\midrule
Atomic transition linewidth & $476(13)$~kHz \\
Optical input power & $58.4(3)$~mW\\
Mixer gain (g) & 0.58(4)\\
Beam waist radius & $1.05(2)$~ mm \\
\toprule
\multicolumn{2}{l}{\textbf{Amplifier Seed: RIO Planex Laser}}\\
\midrule
Instantaneous linewidth & $ 1.10(15)$~kHz\\
Error signal slope (p) & $1.0(2)\mu$V/Hz\\
Voltage spectral density ($S_v$) & $1.0(4) \times 10^{-10}$~V$^2$/Hz\\
Power spectral density\footnotemark[1]& $373(22)$Hz$^2$/Hz\\
$\sigma^{(IM)}_y(\tau = 1~s)$& $5.0(2)\times 10^{-14}$\\
$\sigma^{(SN)}_y(\tau = 1~s)$& $7.5(20)\times 10^{-15}$\\
\toprule
\multicolumn{2}{l}{\textbf{Amplifier Seed: SBS Laser}}\\
\midrule
Instantaneous linewidth & $ 8(2)$Hz \\
Error signal slope (p) & $1.08(3)~\mu$V/Hz\\
Voltage spectral density ($S_v$) & $3.6(2) \times 10^{-11}$V$^2$/Hz\\
Power spectral density\footnotemark[1]& $1.8(2)$Hz$^2$/Hz\\
$\sigma^{(IM)}_y(\tau = 1~s)$& $3.46(20)\times 10^{-15}$\\
$\sigma^{(SN)}_y(\tau = 1~s)$& $4.2 (3)\times 10^{-15}$\\
\botrule
\end{tabular}
\footnotetext[1]{At 202 kHz}
%\end{ruledtabular}
\end{table}

%\begin{equation}
%    \nu_{inst}=\pi PSD_{white}
%\end{equation}

\begin{figure}
\includegraphics[width=0.95\textwidth]{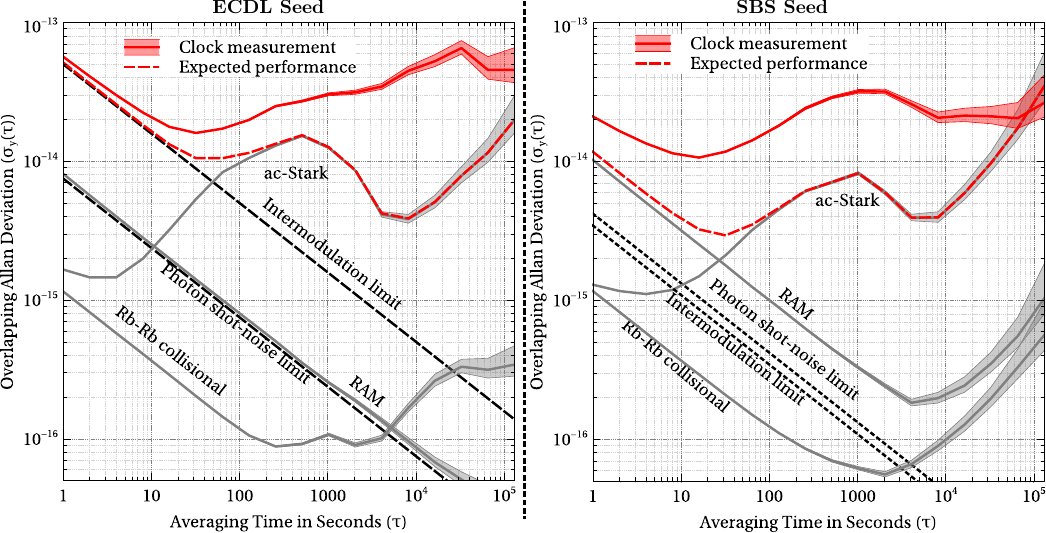}% Here is how to import EPS art
\caption{\label{fig:results}Alongside the measured and expected clock performance are the measured environmental driven instabilities: Rb-Rb collisional, RAM, and ac-Stark all properly scaled to account for the impact on clock stability.}
\end{figure}

To calculate ac-Stark limited performance, the rear-going photodiode measurement was calibrated converting detected voltage to optical power.  This optical power measurement, free-space beam waist, and the ac-Stark transfer function found in \cite{Martin2019} of $2.5\times 10^{-13}$/(mW/mm$^2)$ are used to calculate $\sigma_y^{(ac-Stark)}(\tau)$.  For Rb-Rb collisional limited clock performance out-of-loop temperature measurements scaled by the temperature transfer function in \cite{Martin2018} of $1.1\times 10^{-12}$/K are used to calculate $\sigma_y^{(Rb-Rb)}(\tau)$.  Additionally, prior to clock measurement, a low frequency sinusoid was applied to the dc port of the bias tee whose output is sent to the EOM seen in Figure \ref{fig:experimental_atoms} to determine the proper scaling of detected RAM to clock frequency and calculate RAM limited clock performance, $\sigma_y^{(RAM)}(\tau)$. Finally, shot-noise limit and intermodulation limit using Equations \ref{eq:SN1}, \ref{eq:SN2}, and \ref{eq:IM2} with values measured and reported in Table \ref{tab:table1} are also calculated.  All of these clock limits are shown in Figure \ref{fig:results}.

The fractional frequencies of the references were measured directly in separate experiments.  The stability of the cavity locked laser was measured via optical heterodyne with two independent cavity lasers of similar performance \cite{Gray1974} and determined to be $3.5(3)\times 10^{-15}$ at one second with a drift of $2.5(5)$~kHz/day driven by the laboratory environment.  The cavity laser performs well in the short-term while the the Vector Atomic Evergreen-30 iodine frequency standard was used for long-term comparisons.   The iodine clock was measured against the reference cavity laser using an optical heterodyne as well as compared with a second Vector Atomic Evergreen iodine standard; it's stability was determined to be $3.1(1)\times 10^{-14} /\sqrt{\tau}$ for 100 seconds with a specified drift rate of $<2$~Hz/day \cite{VAEvergreen}.

Figure \ref{fig:results} shows measured clock performance when compared to two stable references.  For simplicity, only the lesser of the two instabilities are shown as the measured result in Fig. \ref{fig:results}, thereby ignoring instabilites and drifts that occur from imperfect references.  Additional details of individual reference clock heterodyne measurements can be found in supplementary Figure \ref{fig:resultsappend}.  Moreover, equation \ref{eq:adev} is used to calculate the expected clock performance given environmental measurements. This calculation is also shown in Figure \ref{fig:results}, and is in good agreement with measured performance over short and long timescales, within a factor of four for the duration of the experiment.

\clearpage

%Figure \ref{fig:results} shows the measured performance of ORAFS with the RIO seed laser and with the SBS seed laser.  Also shown are the ac-Stark, Rb-Rb collisional, and in-phase RAM limits measured during each experiment scaled by the appropriate frequency noise transfer function from \cite{Martin2018}. Finally,  calculated shot-noise and intermodulation limits (assuming $1/\sqrt{\tau}$ behavior) are shown.  

%The vapor cells, mu-magnetic shield, and heating elements were identical to the systems described in \cite{Lemke2022}.  However, to increase the intensity and reduce the photon shot-noise limit of the frequency standard a different collimator was utilized reducing the beam waist radius from 2.1(3) mm to 1.05(2) mm. Additionally, the cat's eye retro-reflector was replaced with a flat mirror to reduce pointing errors on the collimator mount as described in \cite{Martin2019}.  

For the ECDL clock-based system, short-term instability is set by the intermodulation limit.  The heterodyne measurement between the stabilized clock laser and either the iodine frequency comb or the cavity laser system yield similar results.  For this clock laser both references are more stable than the two-photon clock.  

For the SBS clock laser, a dramatic decrease in the intermodulation limit is predicted.  Additionally, a discrepancy begins to appear between the cavity and iodine heterodynes.  In this case the two-photon clock outperforms the iodine frequency standard at one second averaging times. However, neither the predicted shot-noise or intermodulation limits explain the one second instability. The RAM of the system appears to be limiting clock performance in the short term, measured performance $<2\times$ worse then expected.  Since there is still excess SNR a future experiment could be planned to demodulate using the third harmonic and further reduce the impact of RAM, at the cost of SNR \cite{Lemke2022, Burck_2004}, or by leveraging more advanced RAM mitigation techniques \cite{Schibli2024}.  

At timescales beyond 100 seconds both clock laser choices show similar performance. The long-term stability (in this case $>1\times10^{5}$~seconds) is limited by power instability via ac-Stark shift.  This could be improved with new ac-Stark mitigation techniques \cite{andeweg2025activecompensationacstark} and using digital servos to reduce voltage setpoint drifts \cite{Lemke2022}.

Mid-term performance limitation remains a bit of a mystery.  The biggest disagreement between measured and expected performance occurs around $10000$~seconds.  Laboratory temperature was oscillating for the duration of each experiment with a period of 2000~seconds.  This oscillating room temperature led to variations in thermal load on the recirculating chiller resistor and materialized as temporary disturbances in stable base-plate temperature. This is the suspected cause for the bump at this timescale of both the expected and measured performance and the disagreement between measured and expected performance.  Initially, small changes in optical alignment, driven by temperature fluctuations of the free-space optical breadboard, were suspected as the cause for this mid-term mystery.  Therefore, a silicon quadrant photodiode was placed at the rear-going pick-off location.  Frequency stability data and environmental data were collected as the temperature set-point of the free-space optical breadboard was varied.  The total observed intensity of the light appears to vary under thermal fluctuations on the quadrant photodiode a known problem with silicon detectors \cite{Ferri_2014}, however, the location of the centroid moved under thermal fluctuations, see the supplementary Figure \ref{fig:alignment} for further details.  Small heaters were also installed around each photodiode, the Glan polarizer, the fiber collimator, and the retro-reflector.  While shifts were measured while activating these heaters, no measurable frequency shifts were observed when the collimator heater was energized.  The authors suspect that a combination of temperature gradients causing Rb-Rb pressure shifts, residual temperature dependent photodiode detectivity variations on the InGaAs photodiodes, and alignment ac-Stark shifts are the causes for the both mid-term instability, and the disagreement between expected and measured performance.

%we added heaters on the base-plate at the location of several vital free-space components: the collimator, Glan-Taylor polarizer, flat-mirror retro-reflector, and both photodiodes for RAM and power stabilization/monitor.  Additionally,   

%  

\section{Discussion}

We prioritized improvement of short-term clock stability and have demonstrated that using an SBS laser in combination with increased optical intensities can reduce the short-term instability of this two-photon transition in Rb.  We were successful in increasing the short-term stability by over a magnitude over our previous reports.  However, this came at the cost of increased ac-Stark shift.  Additionally, we discovered that RAM eventually becomes our largest instability in the short-term.  There exists methods of suppressing both RAM \cite{Schibli2024, Burck_2004, Lemke2022} and ac-Stark \cite{andeweg2025activecompensationacstark} in the literature and this frequency standard, in it's current form, would benefit from these mitigating techniques.  Moreover, ovenization of the free-space optical components would yield improvement in mid-term performance and is required for operation in the field. The results obtained here will be important for the development of a deployable optical atomic clocks with high frequency stability across all timescales.  Deployable optical clocks are vital component that will become essential for future communication, navigation and sensing technologies.


%\bibliography{clock_papers.bbl}% Produces the bibliography via BibTeX.

%\printbibliography

    \section*{Acknowledgments}
We would like to thank Myles Silfies for many useful discussions.  Any mention of commercial products within this article is for information only; it does not imply recommendation or endorsement by the Department of the Air Force, the Department of War, or the U.S. Government.  Approved for public release, distribution is unlimited. Public Affairs release approval \# AFRL20260505 The views expressed are those of the authors and do not reflect the official policy or position of the Department of the Air Force, the Department of War, or the U.S. Government.

    \section*{Author Contributions}
    K.W.M., K.C, S.K, and S.E.E. and R.B. designed the experiment. K.W.M. took and analyzed the data.  K.W.M. write the first draft. A.I., D.J.B. and K.L. designed and developed the SBS resonator. All authors worked on revisions. K.C, S.K, and S.E.E. provided supervision.   
    \section*{Funding}
    This work was supported in part by the Air Force Research Labs and COSMAIC under award FA9453-20-2-001 and sub-award 282109-873R. 
        \section*{Data availability}
Data underlying the results presented here can be provided from the authors upon reasonable request.

    \section*{Ethics declarations}
    \subsection*{Competing interests}
    The authors declare no competing interests.

\section*{Supplementary information}
%\section{\label{sec:append2} A} 

\begin{figure*}
\includegraphics[width=0.95\textwidth]{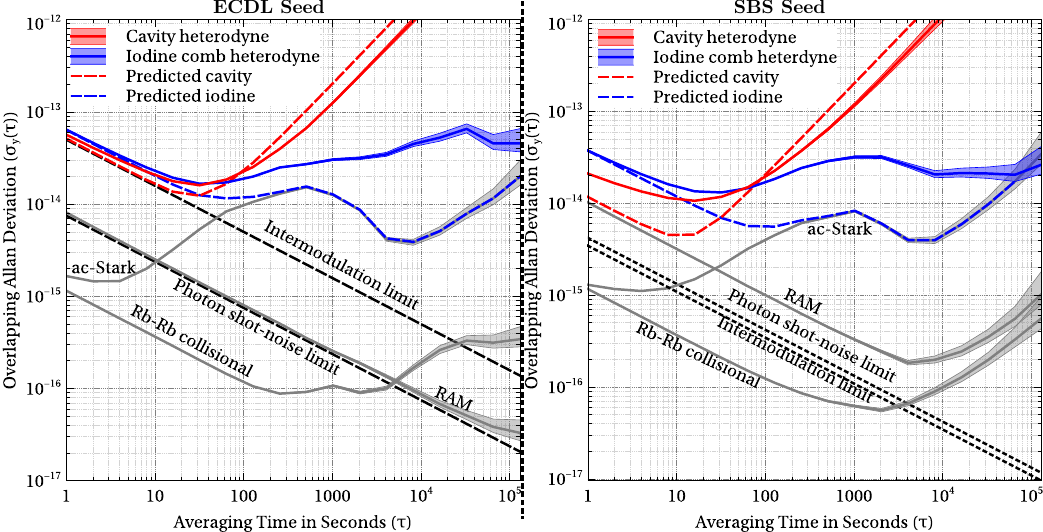}% Here is how to import EPS art
\caption{\label{fig:resultsappend}Solid blue line shows the measured heterodyne between the iodine comb and the two-photon clock, while the dashed blue line show expected performance of this heterodyne including expected reference instabilities and drifts.  Solid red line shows the measured heterodyne between the cavity laser and the two-photon clock, while the dashed red line show expected performance of this heterodyne including expected reference instabilities and drifts.  Alongside the measured and expected clock performance are the measured environmental driven instabilities: Rb-Rb collisional, RAM, and ac-Stark properly scaled to account for their impact on clock shift }
\end{figure*}

%We can calculate the percent change in quad detector power for a beam displacement of $d$, in only the x direction, by evaluating the following, 
%\begin{equation}
%    -P+\frac{4P}{\pi w_0^2}\int_{-\infty}^{\infty} %\int_{0}^{\infty}\bigg[e^{-2\frac{(x-d)^2+y^2}{w_0^2}}\bigg]dxdy,
%\end{equation}
%and normalizing by our optical power $P$.  Evaluating this integral and normalizing by power yields, 
%\begin{equation}
%\erf{\bigg(\frac{\sqrt{2}d}{w_0}\bigg)}.
%\end{equation}
%This equation shows the normalized percent change we can expect for a displacement d on the quadrant detector. For small displacement this can be approximated to first orderto be,
%\begin{equation}
%    \frac{2\sqrt{2} d}{\pi w_0}
%\end{equation}

%\section{\label{sec:append} B} 
%We use equation 6 from \cite{Martin2019} to calculate the spatially weighted average of average of a forward going Gaussian beam and an anti-parallel retro-reflected beam displaced by distance $d$.  This weighted spacial intensity can be written,
%\begin{equation}
%    \bar{I}=\frac{8P}{3\pi w_0^2}\bigg(e^{-\frac{d^2}{3w_0^2}}\bigg),
%\end{equation}
%where $w_0$ is our beam waist radius of 1.05(2)~mm, and $P$ is our incident optical power of 58.4(3)~mW.  Compared to the perfect alignment case, the Gaussian beam center would only have to move by 0.1~mm for a shift of $3\times 10^{-14}$.

%Beam movement is measured and shown with clock frequency changes in Figure \ref{fig:alignment}.  Additionally, changes in PMT temperature could lead to unmeasured thermal gradients at the Rb vapor cells.  

\begin{figure}
\includegraphics[width=\columnwidth]{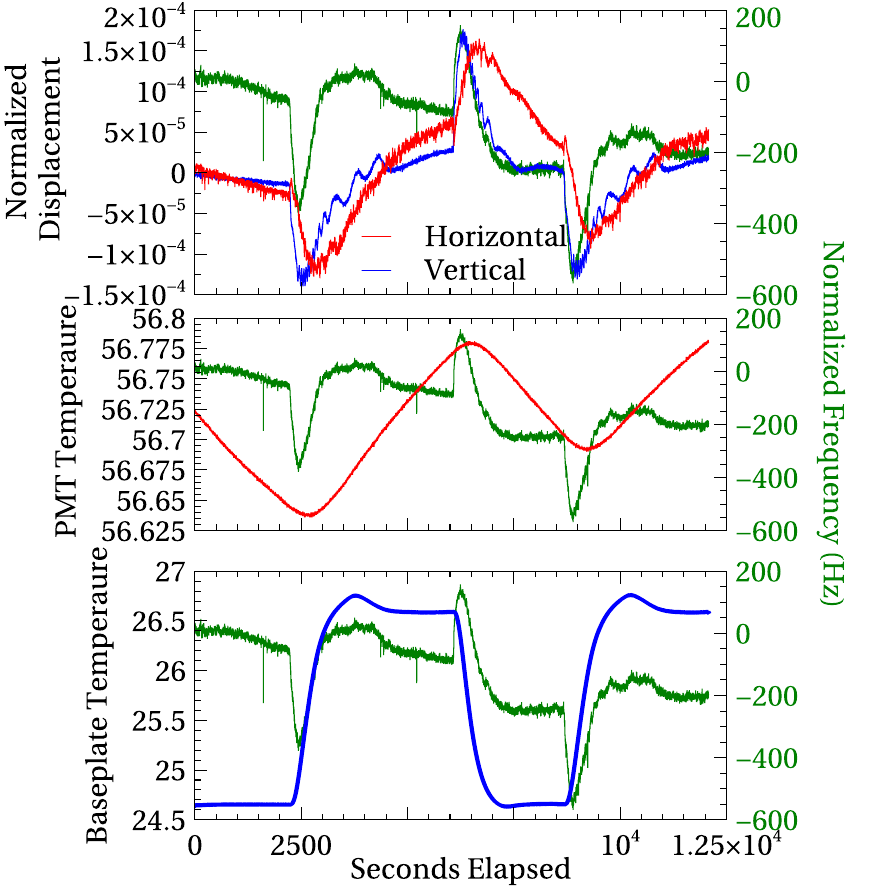}% Here is how to import EPS art
\caption{\label{fig:alignment} The top graph shows the measured vertical (blue) and horizontal (red) alignment changes as the base plate temperature is varied as well as the frequency shifts (green).  The middle graph shows the PMT temperature change (red) and frequency change (green) and the bottom graph shows the baseplate temperature change (blue) along with frequency shifts (green). }
\end{figure}

\end{document}